\documentclass[pdflatex,sn-nature]{sn-jnl}

\usepackage{graphicx}%
\usepackage{multirow}%
\usepackage{amsmath,amssymb,amsfonts}%
\usepackage{amsthm}%
\usepackage{mathrsfs}%
\usepackage[title]{appendix}%
\usepackage{xcolor}%
\usepackage{textcomp}%
\usepackage{manyfoot}%
\usepackage{booktabs}%
\usepackage{algorithm}%
\usepackage{algorithmicx}%
\usepackage{algpseudocode}%
\usepackage{listings}%

\usepackage{lineno}

\theoremstyle{thmstyleone}%
\theoremstyle{thmstyletwo}%

\theoremstyle{thmstylethree}%

\begin{document}
 

\title[Generative AI for subgrid turbulence in large-eddy simulations]{Generative AI for subgrid turbulence in large-eddy simulations}


\author*[1]{\fnm{Yu} \sur{Cheng}}\email{yucheng1@g.harvard.edu}

\author[2]{\fnm{Tianle} \sur{Liu}}
 




\affil[1]{\orgdiv{Department of Earth and Planetary Sciences}, \orgname{Harvard University}, \orgaddress{\street{20 Oxford Street}, \city{Cambridge}, \postcode{02138}, \state{MA}, \country{USA}}}

\affil[2]{\orgdiv{Department of Statistics}, \orgname{Harvard University}, \orgaddress{\street{1 Oxford Street}, \city{Cambridge}, \postcode{02138}, \state{MA}, \country{USA}}}



\abstract{
Turbulence governs the transport of momentum, energy, and scalars in many geophysical and engineering flows. In large-eddy simulations (LES), parameterizing subgrid-scale (SGS) stresses remains a central challenge, as unresolved physical processes strongly influence turbulent transport. Traditional SGS models, such as the Smagorinsky-type models and deep neural networks (DNNs), are deterministic and cannot capture the stochastic nature of turbulence. 
Despite its wide application in computer vision and natural language processing, generative artificial intelligence (AI) has not previously been applied to directly compute SGS stresses in three-dimensional turbulent boundary layers at high Reynolds numbers. 
Here we introduce a denoising diffusion probabilistic model (DDPM) to reconstruct SGS stresses from coarse-grained velocity fields in direct numerical simulations of the atmospheric boundary layer. 
The DDPM consistently outperforms Smagorinsky-type models and previous deep neural networks in terms of spatial correlations and probability  distributions for deviatoric stresses, and can be applied to unseen convective stability conditions and resolutions. 
By learning conditional distributions rather than pointwise values, this generative approach opens a new direction for SGS turbulence modeling at high Reynolds numbers. 
 }




\maketitle

\section*{Introduction}\label{sec1}

 Turbulence in the atmospheric boundary layer (ABL, roughly the lowest 1 kilometer of the atmosphere), governs surface–atmosphere exchanges of momentum, heat, moisture, and CO$_2$ that are critical for weather and climate prediction. 
In the ABL, the wide range of flow scales cannnot be resolved by direct numerical simulations (DNS) with current computational resources \cite{pope2000turbulent,lee2015direct,yamamoto2018numerical}, due to high Reynolds numbers \cite{kunkel2006study,cheng2020model,cheng2021logarithmic,cheng2023logarithmic}. 
Large-eddy simulations (LESs) \cite{smagorinsky1963general,lilly1967representation,deardorff1970numerical} only resolve the spatially filtered large eddies, while parameterizing the effects of unresolved subgrid-scale (SGS) turbulence as a function of the resolved flow. 
However, the performance of LES critically depends on the accuracy of the SGS model \cite{germano1991dynamic,bou2005scale,cheng2022deep}.

Widely used SGS stress parameterizations are typically deterministic, e.g., the ones based on the Smagorinsky model \cite{smagorinsky1963general,bardina1980improved,germano1991dynamic,bou2005scale} and deep neural networks (DNNs) \cite{gamahara2017searching,beck2019deep,cheng2022deep,guan2022stable,liu2022investigation,connolly2025deep}. 
The Smagorinsky model and its variants \cite{smagorinsky1963general,bardina1980improved,germano1991dynamic,bou2005scale}, rely on the eddy-viscosity hypothesis. 
These approaches are justified only when there is a clear scale separation between the resolved and unresolved motions, an assumption that often breaks down in real turbulent flows \cite{schmitt2007boussinesq}. 
Many machine learning models, such as DNNs \cite{gamahara2017searching,cheng2022deep} and convolution neural networks \cite{beck2019deep,guan2022stable,liu2022investigation}, do not rely on the eddy-viscosity assumption and have been shown to outperform traditional models in reconstructing SGS stresses in neutral and convective boundary layers \cite{gamahara2017searching,cheng2022deep,jalaali2025multiscale}.

Including a stochastic term in SGS models may help capture the backscatter of energy from unresolved to resolved scales \cite{piomelli1991subgrid,hartel1994subgrid} and improve both resolved and SGS statistics in turbulent channel flows \cite{mason1992stochastic,marstorp2007stochastic,rasam2014stochastic}. 
However, such formulations often suffered from lack of physical consistency, ad hoc parameter tuning, and limited generalizability beyond the specific conditions for which they were calibrated \cite{piomelli1991subgrid,mason1992stochastic,marstorp2007stochastic,rasam2014stochastic}.

Recently, generative artificial intelligence (AI), which is inherently stochastic, has been successfully applied to computer vision \cite{dhariwal2021diffusion,rombach2022high}, natural language processing  \cite{brown2020language,touvron2023llama}, and molecular modeling \cite{jumper2021highly}. 
It has also been applied to climate parameterizations \cite{behrens2025simulating}, idealized ocean modeling \cite{perezhogin2023generative}, super-resolution reconstruction of climate \cite{stengel2020adversarial,hess2022physically,lockwood2024generative} and turbulence \cite{deng2019super,fukami2019super,liu2020deep,subramaniam2020turbulence,fukami2021machine,bode2021using,kim2021unsupervised,yousif2021high,sardar2024spectrally,sofos2025review}. 
It is worth noting that turbulence super-resolution using generative adversarial networks (GANs) has sometimes been described as ``SGS modeling'' \cite{bode2021using,nista2025parallel,maejima2025unsupervised,heyder2024generative}. 
However, such approaches reconstruct fine-scale velocity fields rather than directly predicting SGS stresses, and thus fundamentally differ from the conventional SGS parameterization \cite{smagorinsky1963general,lilly1967representation,deardorff1970numerical}. 
In addition, generative AI has been applied to SGS parameterizations in simplified systems, such Burgers' equation \cite{alcala2021subgrid} and two-dimensional Navier–Stokes equations \cite{dong2025data}, but these applications remain far from the challenges of three-dimensional turbulent boundary layers at high Reynolds numbers.

\begin{figure}
	\centering
	\includegraphics[width=8cm,clip=true, trim = 0mm 0mm 0mm 0mm]{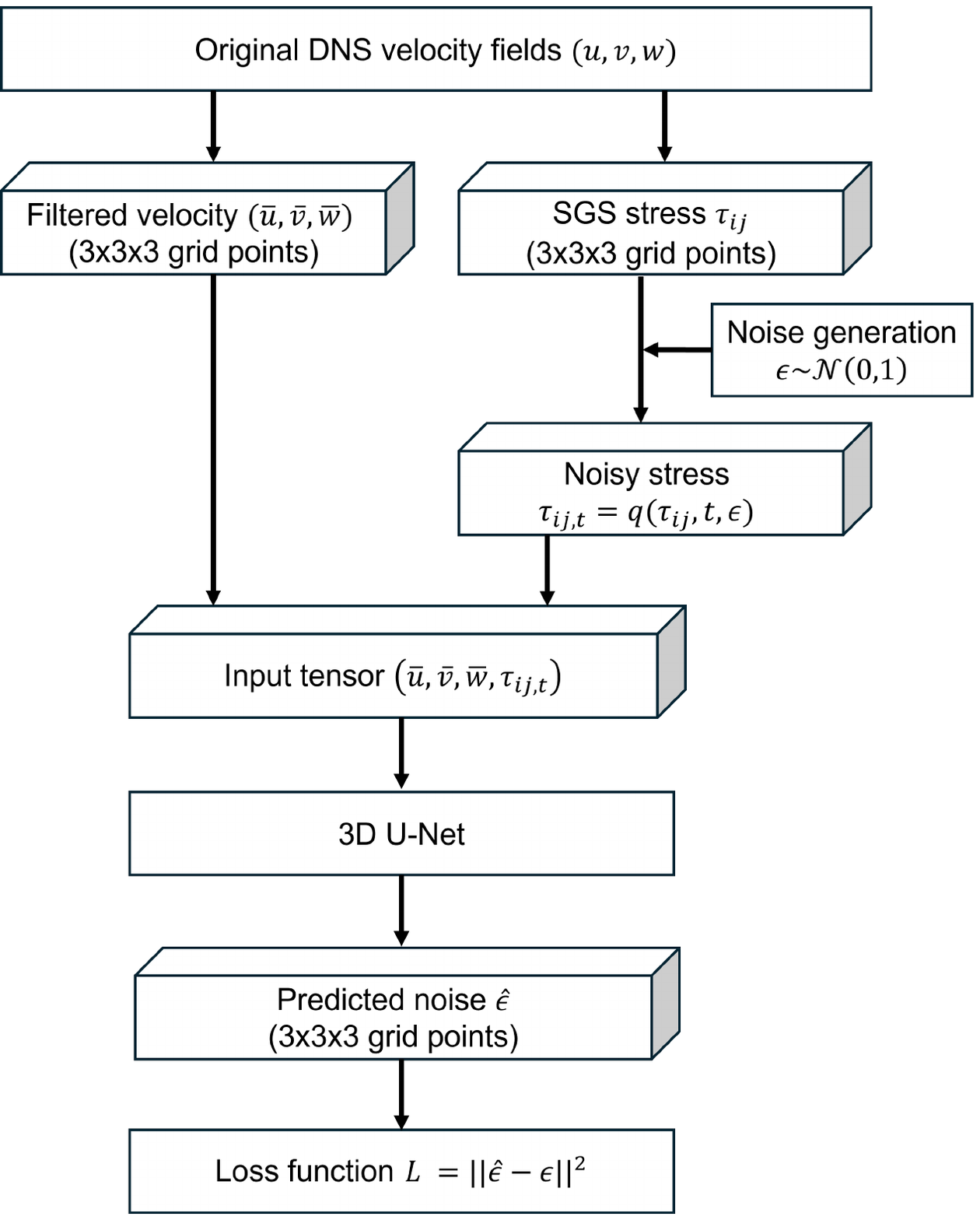}
	    \caption{%
    Schematic of the denoising diffusion probabilistic model (DDPM) used in this study.
    Original DNS velocity fields $(u,v,w)$ are filtered into coarse-grained $(\overline{u},\overline{v},\overline{w})$. 
    Gaussian noise $\epsilon \sim \mathcal{N}(0,1)$ is added to the SGS stress to generate a noisy stress $\tau_{ij,t}$, where $t$ is timestep. 
    The input tensor consists of the filtered velocity components and $\tau_{ij,t}$ over $3\times3\times3$ grid points and is passed through a shallow 3D U-Net \cite{ronneberger2015u,cciccek20163d}. 
    The network predicts the noise $\hat{\epsilon}$. The training minimizes the mean squared error (MSE) loss $L = \|\hat{\epsilon}-\epsilon \|^2$.
    }
 
	\label{fig:schematic_DDPM}
\end{figure}

Here we compute SGS stresses in the convective ABL using a generative AI framework, the denoising diffusion probabilistic model (DDPM) \cite{sohl2015deep,ho2020denoising,nichol2021improved}. 
The DDPM is trained to learn the conditional distribution of SGS stresses from local velocity fields, gradually transforming random noise into physically realistic SGS stresses (Fig. \ref{fig:schematic_DDPM}). 
The training data are coarse-grained DNSs of the ABL ranging from highly convective to weakly convective conditions \cite{li2018implications,cheng2021logarithmic,cheng2022deep}. 
The model is evaluated offline (\textit{a priori}) by reconstructing SGS stresses from unseen coarse-grained DNS velocity fields. 
Across convective stability conditions and spatial resolutions, the SGS stresses from the DDPM maintain high correlations with DNS stresses and preserve the probability distributions of the latter. 
Although online (\textit{a posteriori}) LES tests are left for future work, this study provides a benchmark for applying generative AI to SGS stress modeling and demonstrates its potential to advance three-dimensional turbulence parameterizations at high Reynolds numbers.

\section*{Results}

\subsection*{DDPM captures SGS stresses at unseen convective stability conditions}

We evaluate the capability of the trained DDPM to reconstruct SGS stresses in convective boundary layers spanning a wide range of stability conditions. 
The DDPM\_multiSh model is trained on both highly convective (Sh2, $z_i/L = -678.2$) and weakly convective (Sh20, $z_i/L = -7.1$) cases, and tested on an unseen moderately convective (Sh5, $z_i/L = -105.1$) case (see Materials and Methods for more information on the DNS data \cite{li2018climate,cheng2021logarithmic,cheng2022deep}), where $z_i$ is the boundary layer height and $L$ the Obukhov length \cite{obukhov1946turbulence}.

\begin{figure}
	\centering
	\includegraphics[width=8cm,clip=true, trim = 0mm 0mm 0mm 0mm]{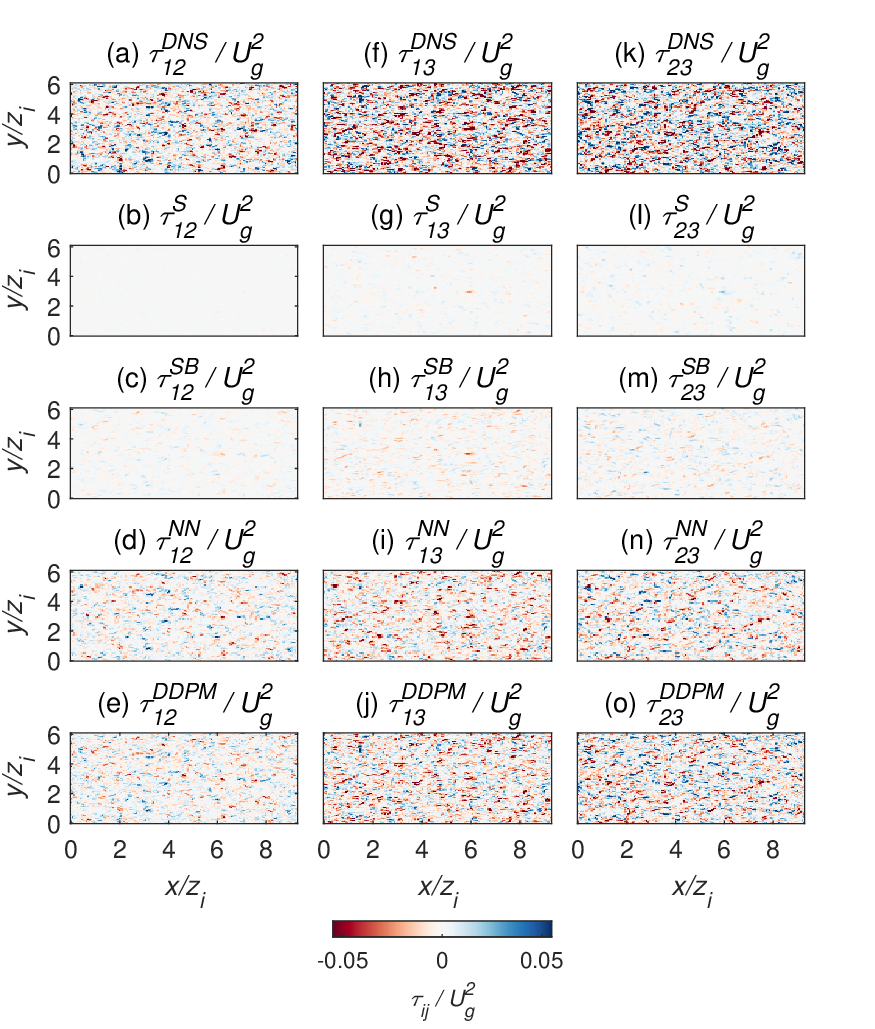}
	\caption{SGS stresses of $x$--$y$ plane normalized by $U_g^2$ at 
$z^+ = \dfrac{z}{(\nu/u_\tau)} = 58.2$ in the log-law region \cite{cheng2021logarithmic} of the moderately convective dataset Sh5. $U_g$ is geostrophic wind, $z_i$ is boundary layer height, 
$\nu$ is kinematic viscosity and $u_\tau$ is friction velocity. 
From top to bottom: SGS stresses calculated from DNS data ($\tau_{ij}^{\mathrm{DNS}}$), the Smagorinsky model ($\tau_{ij}^{\mathrm{S}}$), 
the Smagorinsky--Bardina mixed model  ($\tau_{ij}^{\mathrm{SB}}$), the DNN model from Cheng et al. \cite{cheng2022deep}  ($\tau_{ij}^{\mathrm{NN}}$), and the proposed DDPM\_multiSh ($\tau_{ij}^{\mathrm{DDPM}}$). Here the DDPM\_multiSh is trained on the highly convective (Sh2) and weakly convective (Sh20) cases, and tested on the unseen moderately convective (Sh5) case. } 
 
	\label{fig:DDPM_tau_2Dcontour}
\end{figure}

The predicted deviatoric stresses from the DDPM ($\tau_{ij}^{\mathrm{DDPM}}$) are compared with the SGS stresses from DNS ($\tau_{ij}^{\mathrm{DNS}}$), the Smagorinsky model ($\tau_{ij}^{\mathrm{S}}$) \cite{smagorinsky1963general}, the Smagorinsky–Bardina mixed model ($\tau_{ij}^{\mathrm{SB}}$) \cite{bardina1980improved}, and the DNN model ($\tau_{ij}^{\mathrm{NN}}$) \cite{cheng2022deep} in the $x$--$y$ plane at a representative height in the logarithmic region (Fig. \ref{fig:DDPM_tau_2Dcontour}) \cite{cheng2021logarithmic}, which is equivalent to the log-law region in turbulent channel flows \cite{karman1930mechanische,marusic2013logarithmic}. 
The DDPM prediction closely reproduces the spatial distribution of $\tau_{ij}^{\mathrm{DNS}}$, especially the local fluctuations, with a performance comparable to the DNN model \cite{cheng2022deep}. 
In contrast, both $\tau_{ij}^{\mathrm{S}}$ and $\tau_{ij}^{\mathrm{SB}}$ fail to capture the large magnitude of spatial variations of $\tau_{ij}^{\mathrm{DNS}}$.

\begin{figure}
	\centering
	\includegraphics[width=7 cm,clip=true, trim = 00mm 0mm 00mm 0mm]{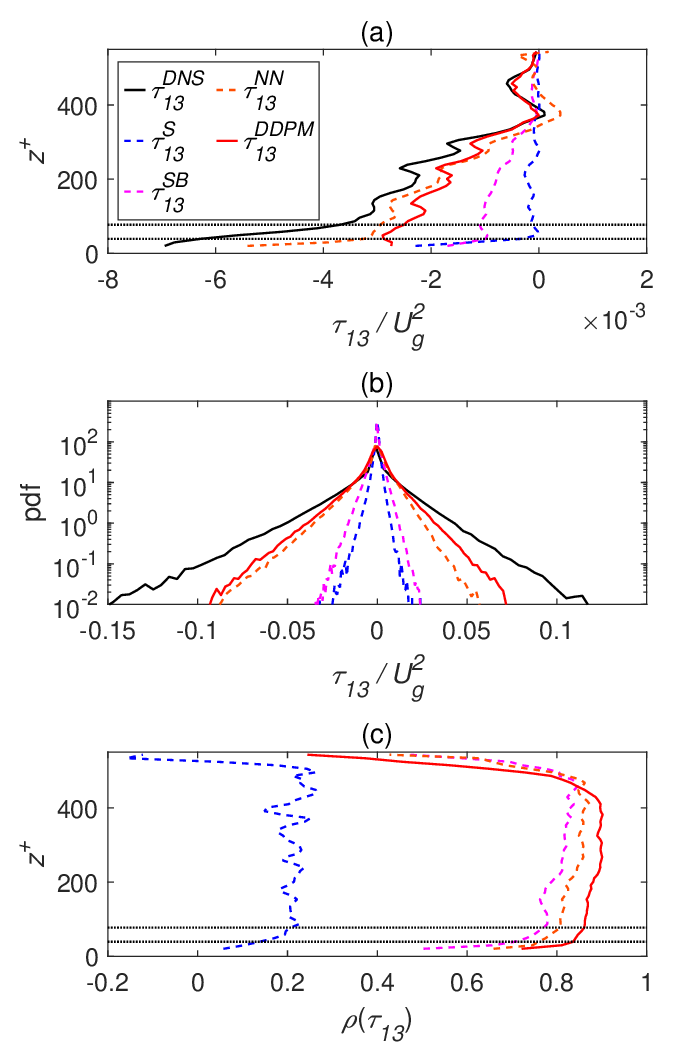}
	\caption{Comparison of $\tau_{13}^\mathrm{DNS}$ from DNS data, $\tau_{13}^\mathrm{S}$ from the Smagorinsky model, 
$\tau_{13}^\mathrm{SB}$ from the Smagorinsky-Bardina model, $\tau_{13}^\mathrm{NN}$ from the DNN model \cite{cheng2022deep}, and $\tau_{13}^\mathrm{DDPM}$ from DDPM\_multiSh. 
(a) Mean SGS stresses of $x$--$y$ plane at different height $z^+$. 
(b) Probability distribution function (pdf) of SGS stresses in the whole DNS field. 
(c) The correlation ($\rho$) between $\tau_{13}^{\mathrm{DNS}}$ and the predicted $\tau_{13}$ from each model in the $x$--$y$ plane at different height $z^+$. The analyzed dataset is the moderately convective Sh5, while the DDPM\_multiSh is trained on the highly convective (Sh2) and weakly convective (Sh20) cases. 
$U_g$ is geostrophic wind, $\rho$ is correlation, and $z^+$ is the normalized distance to the wall. 
In (a) and (c), the height range $z^+$ between the two dotted lines denotes the log-law region \cite{cheng2021logarithmic} in convective boundary layers.}
	
	\label{fig:DDPM_tau_mean_pdf}
\end{figure}

We further compare the vertical profiles of horizontally averaged $\tau_{13}$ from different SGS models (Fig. \ref{fig:DDPM_tau_mean_pdf}a). 
The DDPM captures the magnitude and vertical variations of $\tau_{13}^{\mathrm{DNS}}$, particularly in the logarithmic layer \cite{cheng2021logarithmic} and mixed layer, with an accuracy close to that of the DNN model \cite{cheng2022deep}. 
This is promising because the DDPM learns the distribution of SGS stresses, whereas the DNN model is designed to learn the pointwise values. 
In contrast, the Smagorinsky and Smagorinsky-Bardina models systematically underestimate the magnitude of $\tau_{13}$.  
In addition, the corresponding probability density functions (PDFs) of $\tau_{13}^{\mathrm{DDPM}}$ better match the broader tails of the $\tau_{13}^{\mathrm{DNS}}$ PDFs (Fig. \ref{fig:DDPM_tau_mean_pdf}b) than the other three SGS stresses. 
In particular, the PDFs of $\tau_{13}^{\mathrm{S}}$ and $\tau_{13}^{\mathrm{SB}}$ are too centered near zero, consistent with the small spatial fluctuations in the $x-y$ plane (Fig. \ref{fig:DDPM_tau_2Dcontour}).

The correlation between predicted and true SGS stresses is the most widely used metric in offline (\textit{a priori}) evaluations of SGS models \cite{clark1979evaluation,meneveau2000scale,gamahara2017searching,beck2019deep,cheng2022deep}. 
The correlation ($\rho$) between $\tau_{13}^{\mathrm{DNS}}$ and the predicted $\tau_{13}$ from each model is shown as a function of height (Fig. \ref{fig:DDPM_tau_mean_pdf}c). 
In the mixed layer above the log-law region, the correlation between  $\tau_{13}^{\mathrm{DNS}}$ and $\tau_{13}^{\mathrm{S}}$ remains around 0.2, consistent with previous studies \cite{clark1979evaluation,liu1994properties}. 
The DDPM produces correlation values exceeding 0.85 throughout the mixed layer. 
The correlation from $\tau_{13}^{\mathrm{DDPM}}$ in the middle mixed layer and the log-law region is about 0.05 higher than that from $\tau_{13}^{\mathrm{NN}}$, and about 0.1 higher than that from $\tau_{13}^{\mathrm{SB}}$ \cite{bardina1980improved}.

Over the entire DNS domain, the correlation between $\tau_{13}^{\mathrm{DDPM}}$ and $\tau_{13}^{\mathrm{DNS}}$ remains above 0.83 across all convective stability conditions and spatial resolutions (Fig.~\ref{fig:DDPM_tau13_multiC_multiT_box555}), exceeding the correlations previously reported for convective boundary layers \cite{cheng2022deep}. 
When compared to other turbulent flows, the correlation from DDPM\_multiSh in our convective boundary layers is higher than that obtained in most earlier SGS studies of isotropic or channel flows \cite{clark1979evaluation,bardina1980improved,gamahara2017searching,beck2019deep}. 
Besides, DDPM\_multiSh\_7$\times$7$\times$7 (trained on a patch size of 7$\times$7$\times$7) produces correlations up to 0.95 (Fig.~\ref{fig:DDPM_tau13_multiC_multiT_box555}a), comparable to the values reported by \cite{liu2022investigation} in turbulent channel flows.

When trained on the highly convective (Sh2, $z_i/L = -678.2$) and weakly convective (Sh20, $z_i/L = -7.1$) cases, the DDPM\_multiSh predicts deviatoric SGS stresses for the moderately convective (Sh5, $z_i/L = -105.1$) case better than the Smagorinsky model \cite{smagorinsky1963general}, the Smagorinsky-Bardina mixed model \cite{bardina1980improved}, and the DNN model \cite{cheng2022deep} in terms of correlation and probability density distributions. 
Thus, the DDPM learns the spatial distribution of SGS stresses across a range of convective stability conditions.

\subsection*{Patch-size tradeoff between accuracy and efficiency}

\begin{figure}
	\centering
	\includegraphics[width=8cm,clip=true, trim = 00mm 0mm 00mm 0mm]{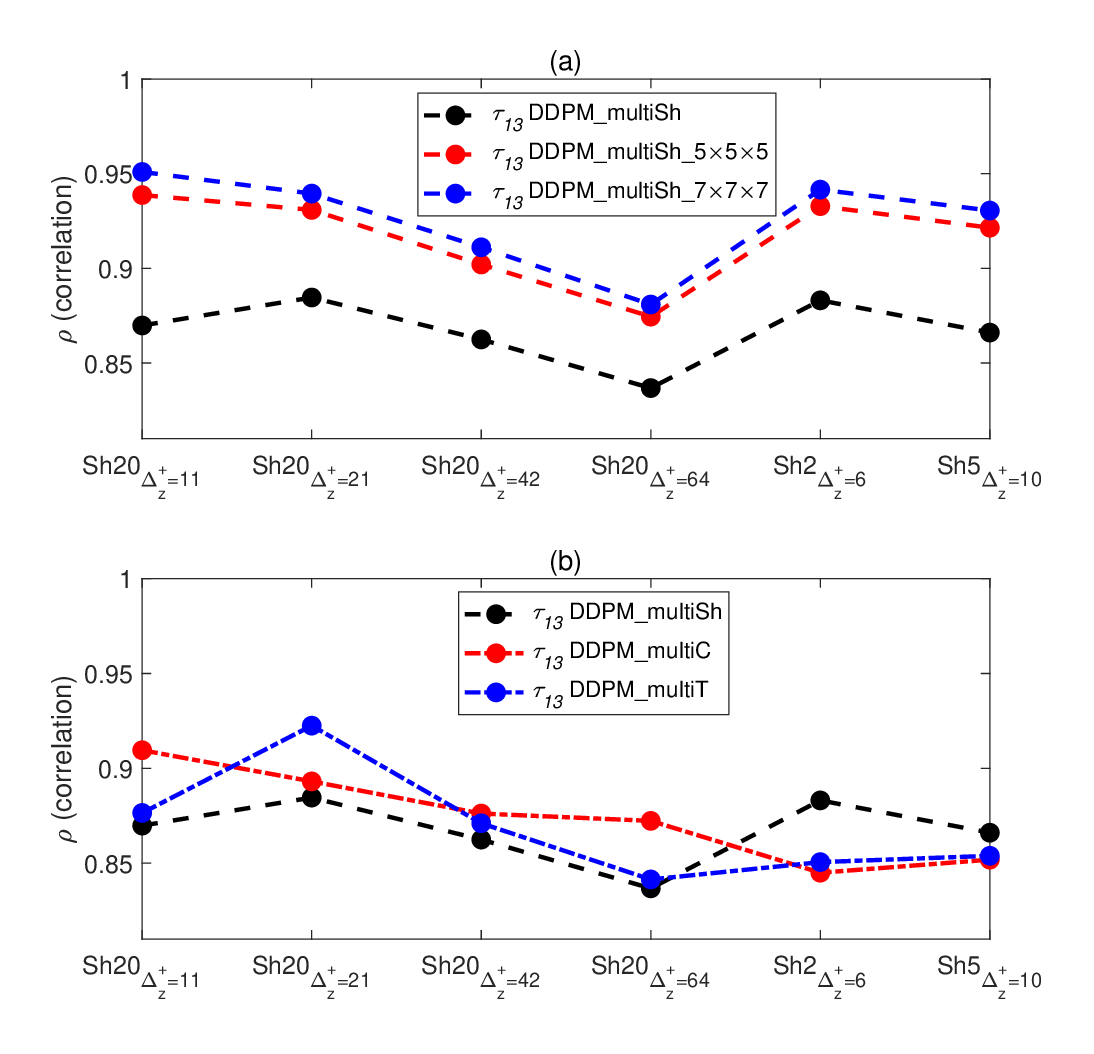}
\caption{Comparison of the correlation coefficients ($\rho$) between $\tau_{13}^{\mathrm{DNS}}$ and $\tau_{13}^{\mathrm{DDPM}}$ from different DDPM over the whole DNS field. (a) Comparison of DDPM\_multiSh (trained on Sh20 at $\Delta_z^+=21$ and Sh2 at $\Delta_z^+=6$), DDPM\_multiSh\_5$\times$5$\times$5 and DDPM\_multiSh\_7$\times$7$\times$7. The DDPM\_multiSh is trained on velocity and SGS stress patches of size 3$\times$3$\times$3, while the latter two models are trained on patches of size 5$\times$5$\times$5 and 7$\times$7$\times$7, respectively. 
(b) Comparison of DDPM\_multiSh, DDPM\_multiC (trained on Sh20 at $\Delta_z^+=11$, 21 42 and 64), and DDPM\_multiT (trained on Sh20 at $\Delta_z^+=21$ across multiple time steps). None of the datasets shown along the $x$-axis are used during the DDPM training. For example, DDPM\_multiT (trained on Sh20 at $\Delta_z^+=21$) is evaluated on different, unseen time steps. }

	\label{fig:DDPM_tau13_multiC_multiT_box555}
\end{figure}

We tested different training patch sizes (Fig. \ref{fig:DDPM_tau13_multiC_multiT_box555}a), where the patch size refers to the local cube of the coarse-grained velocity components ($\overline{u},\overline{v},\overline{w}$) and noisy SGS stresses. 
Increasing the patch size from $3 \times 3 \times 3$ to $5 \times 5 \times 5$ or $7 \times 7 \times 7$ leads to a modest improvement of over 0.03 in the correlation between $\tau_{13}^{\mathrm{DNS}}$ and $\tau_{13}^{\mathrm{DDPM}}$. 
Thus, SGS stresses are mainly determined by the immediate neighboring velocity fields, consistent with the findings from the previous DNN model \cite{cheng2022deep}. 
As a result, we selected $3 \times 3 \times 3$ patches for the final model (DDPM\_multiSh) as a tradeoff between predictive accuracy and computational efficiency.

\subsection*{Convective stability-awareness of DDPM}

To test if the initial training data must span various convective conditions to capture coherent structures in the ABL \cite{agee1973review,atkinson1996mesoscale,li2011coherent,salesky2017nature}, 
we also train DDPM\_multiT on the weakly convective Sh20$_{\Delta_z^+=21}$ ($z_i/L = -7.1$) at different time steps.

For the weakly convective Sh20 ($z_i/L = -7.1$) at vertical resolutions of $\Delta_z^+=11$, 42, and 64, DDPM\_multiT produces a little higher correlation (around 0.01) for $\tau_{13}$, compared to DDPM\_multiSh (Fig. \ref{fig:DDPM_tau13_multiC_multiT_box555}b).
This can be explained by the fact that the former model is trained on the weakly convective Sh20, while the latter model is trained on both the weakly convective Sh20 and the highly convective Sh2 ($z_i/L = -678.2$). 
For the Sh20 dataset at a vertical resolution of $\Delta_z^+=21$, the difference in correlation between DDPM\_multiT and DDPM\_multiSh is around 0.04, which is higher than the differences in Sh20 at other resolutions ($\Delta_z^+=11$, 42 and 64). 
This is likely caused by the fact that DDPM\_multiT is trained on Sh20 at $\Delta_z^+=21$ and better predicts SGS stresses at this spatial resolution compared to other resolutions.

For the moderately convective Sh5 ($z_i/L = -105.1$) and highly convective Sh2 ($z_i/L = -678.2$), DDPM\_multiT (trained on the weakly convective Sh20) produces a lower correlation for $\tau_{13}$ (Fig. \ref{fig:DDPM_tau13_multiC_multiT_box555}b) compared with DDPM\_multiSh (trained on both Sh20 and the highly convective Sh2). This is expected as only the latter model is trained on the more convective cases.

If we want to make predictions on datasets with the same convective stability as the training data, it is enough to train on the dataset with one convective stability. 
In contrast, to predict under unseen stability conditions, it is better to train on a range of cases from highly convective to weakly convective conditions that cover the target stability.

\subsection*{Scale-awareness of DDPM}

The DDPM\_multiT trained on Sh20 at $\Delta_z^+=21$ maintains high correlations ($\rho > 0.84$) when applied to Sh20 at $\Delta_z^+=11$, 42 and 64 (Fig. \ref{fig:DDPM_tau13_multiC_multiT_box555}b), demonstrating the ability to generalize across resolutions. 
In addition, DDPM\_multiT produces high correlations ($\rho >= 0.85$) over the finer-resolution Sh5 (at $\Delta_z^+=10$) and Sh2 (at $\Delta_z^+=6$). 
Therefore, DDPM\_multiT trained on datasets with the resolution $\Delta_z^+=21$ generalizes well to unseen datasets that are up to three times finer ($\Delta_z^+=6$) or coarser ($\Delta_z^+=64$).

To test impact of multi-resolution training on DDPM, we apply different coarse-graining filters to Sh20 and train DDPM\_multiC on the resulting vertical resolutions of $\Delta_z^+=11$, 21 42 and 64.    
For Sh20 at the finer resolution $\Delta_z^+=11$ and the coarse resolution $\Delta_z^+=64$, DDPM\_multiT produces a lower correlation (around 0.03) for $\tau_{13}$, compared with DDPM\_multiC (Fig. \ref{fig:DDPM_tau13_multiC_multiT_box555}b).
This is due to fact that the former model is trained only on $\Delta_z^+=21$, while the latter model is trained across multiple resolutions ($\Delta_z^+=11$, 21 42 and 64). 
Therefore, training on a wide range of spatial resolutions leads to better generalization across resolutions than training on a single resolution.

\section*{Discussion}

The proposed DDPM produces higher correlation coefficients and better reproduces the PDFs of deviatoric SGS stresses than the Smagorinsky model \cite{smagorinsky1963general}, the Smagorinsky–Bardina mixed model \cite{bardina1980improved}, and the deterministic DNN model \cite{cheng2022deep}. 
This result is not intuitive, since DDPMs are not trained to directly minimize the error of pointwise SGS stresses.
Instead, they are designed to model the conditional distribution of SGS stresses given the resolved velocity field by learning to reverse a diffusion (noise corruption) process \cite{sohl2015deep,ho2020denoising}. During training, the true SGS stress is not directly fed into the DDPM.
The DDPM actually learns to predict the noise added to the stresses. 
The fact that this indirect, distribution-learning approach performs better than deterministic models opens a new direction for SGS parameterizations in three-dimensional turbulent boundary layers at high Reynolds numbers.

Many traditional SGS models, including the Smagorinsky and deterministic DNNs, implicitly assume that SGS stresses are a deterministic function of local resolved velocity. 
However, SGS stresses are inherently stochastic due to the nature of turbulence as well as the backscatter of energy from unresolved to resolved scales \cite{piomelli1991subgrid,hartel1994subgrid,mason1992stochastic,marstorp2007stochastic,rasam2014stochastic}.   
The same filtered velocity field may correspond to multiple physically plausible fine-scale DNS velocity fields and thus a distribution of possible SGS stresses.  
The improved performance of DDPM in correlations and PDFs may come from the noise sampling during training, which enables the model to capture the distribution of SGS stresses conditioned on the velocity field.  
In addition, the injected noise during training may act as a form of regularization, which could enhance generalization and robustness to noisy labels compared with deterministic models  \cite{ho2020denoising,song2020denoising}.

\section*{Conclusions}

We develop a DDPM to reconstruct SGS turbulence stresses using resolved velocity fields in the ABL at high Reynold numbers. The key findings from this study are summarized below:
 
\begin{itemize}
    \item The DDPM outperforms the Smagorinsky model, the Smagorinsky--Bardina mixed model, and the DNNs in producing higher correlation coefficients and more realistic PDFs for deviatoric stresses. These improvements are observed in both the log-law region and mixed layer of the ABL.

    \item The DDPM predicts accurate SGS stresses using only a 3$\times$3$\times$3 velocity patch, demonstrating that SGS stresses mainly depend on the immediate velocity neighborhood. Larger patch sizes can increase correlation coefficients but also increase computational cost. 
 
     \item The DDPM trained on two extreme stability conditions (i.e., highly convective and weakly convective boundary layers) generalizes well to intermediate convective conditions. This suggests that the model is capable of learning SGS stress distributions that interpolate across the stability conditions. 

     \item The DDPM can be used to predict SGS stresses in datasets with resolutions up to three times finer or coarser than the training data.

\end{itemize}

This study demonstrates the potential of generative AI for SGS stress parameterizations in three-dimensional turbulence boundary layers at high Reynolds numbers. 
Integrating the trained DDPM into LESs for online \textit{a posteriori} tests and incorporating physics-based models into data-driven frameworks will be pursued in future work.

\section*{Methods}

\subsection*{DNS dataset}

We use a suite of high-resolution DNSs of the ABL to train and test the proposed DDPM.  
The incompressible Navier-Stokes equations with Boussinesq approximation are solved to resolve the convective boundary layers \cite{li2018implications,cheng2021logarithmic,cheng2022deep}. 
Numerical details on the DNS code can be found in \cite{shah2014direct}. 
The viscous sublayer is excluded from all DNS datasets as SGS parameterization is not designed for in this region. 
The bottom boundary is no-slip and impermeable, while the top boundary is free-slip and impermeable. 
The temperature field is forced with a constant flux at the bottom surface and a zero-flux condition at the top domain.  
The top 25\% grid points is occupied by a sponge layer in vertical direction to dissipate gravity waves \cite{nieuwstadt1993large,li2018implications}.

The stability condition of convective boundary layers is characterized by $z_i/L$, where $z_i$ is the boundary layer height and $L$ the Obukhov length \cite{obukhov1946turbulence}.
The friction Reynolds number is defined as $Re_\tau = \frac{u_\tau z_i}{\nu}$, where $u_\tau$ is the friction velocity and $\nu$ is the kinematic viscosity. 
The convective DNS data are named Sh2, Sh5, and Sh20, where $\mathrm{Re}_\tau = 309$ ($z_i/L = -678.2$), 554 ($z_i/L = -105.1$), and 1243 ($z_i/L = -7.1$), respectively. 
The dataset Sh2 is resolved on grid points of 1200$\times$800$\times$602, while both Sh5 and Sh20 are resolved on grid points of $nx \times ny \times nz=1200 \times 800 \times 626$ in streamwise ($x$), spanwise ($y$) and vertical ($z$) directions, respectively. 
The geostrophic wind $U_g$ defines the initial mean wind profile. 
Each simulation provides streamwise ($u$), spanwise ($v$), and vertical ($w$) velocity components over the 3-dimensional computational domain at different time steps.
$\Delta_z^+ = \frac{\Delta_z u_\tau}{\nu }$ is the vertical grid resolution $\Delta_z$ normalized by $\frac{\nu}{u_\tau}$.
For example, Sh5${_{\Delta_z^+=10}}$ denotes the data Sh5 with a vertical resolution of $\Delta_z^+=10$.

A logarithmic region has recently been identified in both convective \cite{cheng2021logarithmic} and stably stratified boundary layers \cite{cheng2023logarithmic}, where buoyancy effects modify the slope relative to the universal log law observed in neutral boundary layers and pipe flows at high Reynolds number \cite{karman1930mechanische,marusic2013logarithmic}. 
Thus we hypothesize that the DDPM can be extended to higher Reynolds numbers if it accurately captures the SGS stresses within this log-law region.

The DNS datasets are spatially filtered \cite{leonard1975energy} to generate coarse-grained velocity field using a top-hat kernel \cite{clark1979evaluation}.
Similarly to Clark et al. \cite{clark1979evaluation}, the filtered velocity $\overline{u}_i$ is calculated as
\begin{equation}
\overline{u}_i(x,y,z) = \frac{1}{(2k+1)^3} 
\sum_{x'=x-k}^{x+k} \sum_{y'=y-k}^{y+k} \sum_{z'=z-k}^{z+k} 
u_i(x',y',z'), 
\end{equation}
where $2k + 1$ is the coarse-graining factor, $(x,y,z)$ are the coordinates of the points on the original DNS grid, and $\overline{\phantom{u}}$ denotes the top-hat filtering.
The cutoff grid sizes of the DDPM need not fall in the inertial subrange as required by scale-invariant Smagorinsky models. 
The SGS stress tensor is defined as 
\begin{equation}
    \tau_{ij}  = \overline{u_i u_j} - \overline{u}_i \overline{u}_j.
\end{equation}

\subsection*{DDPM for SGS stress parameterization in three-dimensional turbulent boundary layers}

This section describes the data preprocessing, diffusion formulation, model architecture, training procedure, and sampling and inference.

\subsubsection*{Data Preprocessing}

Both the filtered velocities $\overline{u}_i$ and the SGS stresses $\tau_{ij}$ were normalized to nondimensional form. 
Similarly to \cite{cheng2022deep}, the $\overline{u}_i$ and $\tau_{ij}$ fields were divided into overlapping cubic patches of size $3\times 3\times 3$ with stride 1 in each spatial direction. 
Each training sample contained normalized $\overline{u}$, $\overline{v}$, $\overline{w}$, and one SGS stress component $\tau_{ij}$ (e.g., $\tau_{13}$) over a $3\times 3\times 3$ grid box. 
Approximately 3.9 million such patches from the DNS datasets were used in each DDPM training process.

\subsubsection*{Diffusion formulation}
We adopted the DDPM framework introduced by Ho et al. \cite{ho2020denoising}. 
The forward diffusion process gradually corrupted the clean SGS stress $\tau_{ij,0}$ with Gaussian noise ($\epsilon$) over $T$ timesteps according to
\begin{equation}
q(\tau_{ij,t}\mid\tau_{ij,0}) \;=\; \sqrt{\bar{\alpha}_t}\,\tau_{ij,0} \;+\; \sqrt{1-\bar{\alpha}_t}\,\epsilon,
\quad \epsilon \sim \mathcal{N}(0,I),
\end{equation}
where $\{\alpha_t\}_{t=1}^T$ follows a linear noise schedule $\beta_t \in [10^{-4},\,0.02]$, with $\alpha_t = 1-\beta_t$ and $\bar{\alpha}_t = \prod_{s=1}^{t}\alpha_s$. 
Here $I$ denotes the identity over the tensor shape and $\epsilon$ is i.i.d. across the $3\times3\times3$ patches.

When training, a timestep $t$ is sampled uniformly from $\{1,\dots,T\}$, a Gaussian noise $\epsilon$ is generated, and a noised stress $\tau_{ij,t}$ is generated by $q(\tau_{ij,t} \mid \tau_{ij,0})$. The DDPM receives the concatenated tensor
\begin{equation}
x_{\text{in}} = [\overline{u},\overline{v},\overline{w},\tau_{ij,t}] \in \mathbb{R}^{4\times 3\times 3\times 3},
\end{equation}
together with a normalized time embedding $t/T$. The training objective is to predict the injected noise $\epsilon$ from $(x_{\text{in}},t)$ by minimizing
\begin{equation}
\mathcal{L} = \mathbb{E}_{\tau_{ij,0},t,\epsilon}\, \big\| \epsilon - \hat{\epsilon}_\theta(x_{\text{in}},t) \big\|^2,
\end{equation}
where $\hat{\epsilon}_\theta$ denotes the U-Net prediction \cite{ronneberger2015u,cciccek20163d}.

\subsubsection*{Model architecture}
The network $\hat{\epsilon}_\theta(x_{\text{in}},t)$ is implemented as a shallow three-dimensional U-Net \cite{ronneberger2015u,cciccek20163d} modified for the fixed patch size of $3{\times}3{\times}3$ (Fig.~\ref{fig:schematic_DDPM}). 
The input $x_{\text{in}}$ consists of three velocity components $(\overline{u},\overline{v},\overline{w})$ and the noisy SGS stress $\tau_{ij,t}$, while the timestep embedding serves as diffusion-step conditioning.

The architecture maintains the patch size throughout, since no downsampling or upsampling is applied. It consists of four components:
\begin{itemize}
    \item Encoder: a single $3{\times}3{\times}3$ convolution \cite{lecun1998gradient} mapping the four input channels to a base width of 256, followed by group normalization \cite{wu2018group} and sigmoid-weighted linear unit (SiLU) activation \cite{elfwing2018sigmoid}. 
    \item Bottleneck: two residual blocks \cite{he2016deep}, each containing two $3{\times}3{\times}3$ convolutions with group normalization and SiLU activation, combined with identity skip connections. 
    \item Decoder: a $3{\times}3{\times}3$ convolution with group normalization and SiLU activation.
    \item Output: a final $1{\times}1{\times}1$ convolution mapping features to a single channel corresponding to the predicted noise $\hat{\epsilon}$.
\end{itemize}
The output is a tensor of shape $(1,3,3,3)$ for the noise estimation ($\hat{\epsilon}$) over the SGS stress patch.

\subsubsection*{Training procedure}
The dataset (about 3.9 million patches of $3\times 3\times 3$ grid points for $x_{\text{in}}$) was divided into training (80\%) and validation (20\%) subsets. 
Mini-batches of 256 patches were sampled, and the model was trained for 100 epochs. 
The Adam optimizer \cite{kingma2014adam} with a learning rate of $3\times 10^{-4}$ was used for optimization. 
Automatic mixed precision (AMP) \cite{micikevicius2017mixed} on NVIDIA H200 GPUs was employed to accelerate training. 
Validation loss was computed at the end of each epoch using the same forward diffusion process, evaluated without backpropagation.

\subsubsection*{Sampling and inference}
To generate SGS stresses conditioned on velocity fields, we adopted the denoising diffusion implicit model (DDIM) sampler \cite{song2020denoising}, which defines a deterministic reverse process that produces samples statistically consistent with those of the original DDPM. 
This DDIM sampler does not need to inject stochastic noise at each reverse step, which allows accurate generation with far fewer denoising steps (here $n_{\mathrm{steps}} = 50$) compared with the $T=200$ steps used in training. 
Starting from Gaussian noise $\tau_{ij,T} \sim \mathcal{N}(0,I)$, the model integrates backward through a sequence of evenly spaced timesteps to reconstruct the SGS stresses.

At each step $t$, the DDIM predicts the noise $\hat{\epsilon}_\theta(\tau_{ij,t}, \mathbf{\overline{u}}, t)$ and estimates the clean stress 
\begin{equation}
\hat{\tau}_{ij,0}^{(t)} = \frac{1}{\sqrt{\bar{\alpha}_t}}\left(\tau_{ij,t} - \sqrt{1-\bar{\alpha}_t}\,\hat{\epsilon}_\theta(\tau_{ij,t},\mathbf{\overline{u}},t)\right),
\end{equation}
where $\mathbf{\overline{u}} = [\overline{u},\overline{v},\overline{w}]$ denotes the local velocity patch. 
The next iterate is then computed deterministically as 
\begin{equation}
\tau_{ij,t-1} = \sqrt{\bar{\alpha}_{t-1}}\,\hat{\tau}_{ij,0}^{(t)} + \sqrt{1-\bar{\alpha}_{t-1}}\,\hat{\epsilon}_\theta(\tau_{ij,t},\mathbf{u},t), 
\end{equation}
without an additional noise term. 
This contrasts with the DDPM training, where Gaussian noise is added at every step to train the model to perform denoising. 
This approach follows the deterministic update rule of DDIM, which is widely used in practice. 
A synthetic SGS stress patch $\hat{\tau}_{ij,0}$ is produced conditioned on the local velocity field.

The SGS stresses over the whole DNS domain were obtained patch-wise by sliding a $3{\times}3{\times}3$ window across the domain. 
Patch predictions were accumulated into the full field, and the final value at each grid point was obtained by averaging over all overlapping patches. 
This patch-wise DDIM provides efficient inference over large domains and maintains accurate predictions of SGS stresses.

\backmatter


\noindent \\




\bmhead{Acknowledgements}
We thank Kaighin A. McColl for valuable discussions and Qi Li for sharing the DNS datasets used in this study. Training and computations were performed on the FASRC Cannon cluster supported by the FAS Division of Science Research Computing Group at Harvard University.

\bibliography{AllBib}

\end{document}